\documentclass[superscriptaddress,twocolumn,showpacs,showkeys,amsmath,aps,final]{revtex4}
\usepackage{graphicx}% Include figure files
\usepackage{dcolumn}% Align table columns on decimal point
\usepackage[sort&compress]{natbib}
\usepackage{subfigure}
\usepackage{ifpdf}
\usepackage{bm}

\ifpdf
\usepackage[pdftex,
        colorlinks=true,
        pdftitle={ Modeling of chemical processes in the low pressure capacitive RF discharges in a mixture of Ar/C$_2$H$_2$.},
        pdfauthor={ D.A. Ariskin },
        pdfsubject={},
        pdfkeywords={plasmochemistry, gas discharge},
        baseurl={},
        pdfpagemode=UseNone,
        bookmarksopen=true
        pdfoutput=1
        ]{hyperref}
\fi

\begin{document}

\title{Modeling of chemical processes in the low pressure capacitive RF discharges in a mixture of Ar/C$_2$H$_2$ }
\author{D.A. Ariskin}
\affiliation{Institute of Theoretical and Applied Mechanics,
 Russian Academy of Sciences, Novosibirsk 630090, Russia}
\affiliation{Departement Fysica, Universiteit Antwerpen,
Groenenborgerlaan 171, B-2020 Antwerpen, Belgium}
\author{I.V. Schweigert}
\affiliation{Institute of Theoretical and Applied Mechanics,
 Russian Academy of Sciences, Novosibirsk 630090, Russia}
\affiliation{Departement Fysica, Universiteit Antwerpen,
Groenenborgerlaan 171, B-2020 Antwerpen, Belgium}
\author{A.L. Alexandrov}
\affiliation{Institute of Theoretical and Applied Mechanics,
 Russian Academy of Sciences, Novosibirsk 630090, Russia}
\author{A.~Bogaerts}
\affiliation{Departement Chemie, Universiteit Antwerpen,
Universiteitsplein 1, B-2610 Antwerpen, Belgium}
\author{F.M.~Peeters}
\affiliation{Departement Fysica, Universiteit Antwerpen,
Groenenborgerlaan 171, B-2020 Antwerpen, Belgium}

\date{\today}
\begin{abstract}
We study the properties of a capacitive 13.56 MHz discharge properties with a mixture of Ar/C$_2$H$_2$ taking into account the plasmochemistry and growth of heavy hydrocarbons.
A hybrid model was developed to combine the kinetic description for electron motion and the fluid approach for negative and positive ions transport and plasmochemical processes. A significant change of plasma parameters related to injection of 5.8$\%$ portion of acetylene in argon was observed and analyzed. We found that the electronegativity of the mixture is about 30$\%$. The densities of negatively and positively charged heavy hydrocarbons are sufficiently large to be precursors for the formation of nanoparticles in the discharge volume.
\end{abstract}

%\pacs{52.27.Lw, 52.25.-b}
\maketitle

\section{Introduction}
The gas discharge in hydrocarbon mixtures is widely used for carbon film growth. These thin films are of great interest for a wide range of industrial applications due to their extraordinary material properties \cite{1,2,3}.

The advantage of a capacitively coupled radio frequency (CCRF) discharge is that it can be used for producing non conducting films. Noble gases like argon and neon are often used as main background gases for hydrocarbon mixtures as their presence changes morphology of diamond like carbon films and leads to fewer crystalline defects \cite{noble, noble2}.

Plasmochemical processes taking place in the CCRF discharge in a hydrocarbon mixture result in the creation of reactive and neutral species that in its turn leads to film growth on a wafer and to the formation of nanoparticles in the discharge volume. Nucleation of radicals due to the gas phase reactions essentially diminishes the rate of film growth at the substrate. However it was found that such particles can have their own applications like producing effective catalysts and composite coatings \cite{2,4,5}.

The gas phase reactions and nanoparticle formation in the capacitive 13.56 MHz discharge operating in a mixture of argon and acetylene were intensively studied in recent experiments \cite{6,7}. In these experiments the presence of dust particles induces periodic changes of the discharge properties.
Detailed kinetic simulations of the plasma with movable dust in pure argon \cite{dust_in_argon} and for typical experimental conditions \cite{29} were performed. The results of these simulations allowed us to explain the transition between capacitive and resistive modes of discharge glow in the beginning of the dust growth cycle. In these simulations we used the assumption that the gas mixture did not change and plasmochemical processes were not taken into account as well as nanoparticles formation. However the inclusion of chemical reactions in a gas mixture can seriously impact the discharge properties. There is a set of works devoted to the investigation of plasmochemical processes leading to dust formation for different gas mixtures (see for example with SiH$_4$ \cite{8} and with C$_2$H$_2$ \cite{9}).
The main goal of this study is the investigation of plasmochemical processes in a 75 mTorr Ar/C$_2$H$_2$ CCRF discharge leading to dust particles formation and the influence of these processes on discharge properties in the initial stage of nanoparticle growth.

The main mechanism for dust particle formation is considered to be the growth and agglomeration of hydrocarbon chains. The role of certain radicals and ions (precursors) is very important, they initiate the growth process. Formation of precursors is usually a result of a reaction chain initiated by inelastic  electron-molecule collisions (ionization, dissociation, electron attachment or excitation of the molecule).
The rate of a certain process is defined by the cross section, electron energy distribution function (EEDF) and gas pressure. Therefore the accurate calculation of the electron energy distribution function is very important for modeling plasmochemical processes. Calculating the EEDF is a computationally expensive task especially at low gas pressure when the EEDF is not a local function of the electrical field.

In this work we study the plasma dynamics and plasmochemical processes in a capacitive 13.56 MHz discharge in a Ar/C$_2$H$_2$ mixture for the conditions of Bochum experiments \cite{6,7}. The interelectrode distance is 7 cm, the gas pressure is 75 mTorr, the amplitude of applied voltage is 92 V, the gas inlet is 8 sccm for argon and 0.5 sccm for acetylene. As a base for the chemical processes simulation we took the reaction set from \cite{9}.

The structure of the paper is as follows. The description of the model and a discussion of its applicability are given in Sec. \ref{section:mod_desc}. The numerical results are discussed in Sec. \ref{section:res_desc}. The conclusions are given Sec. \ref{section:con_desc}.

\section{Model description}
\label{section:mod_desc}

\subsection{Model overview}
In our hybrid model the electron dynamics is described with the Boltzmann equation and the motion of different types of ions and radicals is modeled with the fluid approach. From the energy distribution function for electrons we calculate the generation rates of ions and radicals from the background gas. The species considered in our model are shown in Table \ref{table:species}. We took the set of species from \cite{9} for pure acetylene and extend it with species specific for the Ar/C$_2$H$_2$ mixture. Balance equations for neutrals and ions include chemical processes. We take into account 146 reactions (section \ref{section:chemistry}). This system of equations is solved self-consistently with the Poisson equation for the electric field distribution.

\begin{table}[ht]
\caption{The different species included in our simulation}
\centering
\begin{tabular}{c c c}
\hline \hline
Neutrals & Ions & Radicals \\ [0.5ex]
\hline
$Ar$  &  $Ar^+$, $ArH^+$  &  \\
$C_2H_2$  &  $C_2H_2^+$, $C_2H^+$, $C_2^+$  &  $C_2H_3$ \\
$C_4H_2$, $C_6H_2$ & $C_4H_2^+$, $C_6H_2^+$, $C_6H_4^+$& $C_4H_3$, $C_6H_3$ \\
 & $C_4H_3^+$, $C_6H_5^+$  & $C_6H_5$ \\
$C_8H_2$, $C_{10}H_2$ & $C_8H_4^+$, $C_8H_6^+$, $C_{10}H_6^+$ & \\
$C_{12}H_2$  &  $C_{12}H_6^+$  &  $C_{12}H_6$ \\
$H_2$  &  $H_2^+$, $H^+$  &  $H$ \\
 & $C_2H^-$, $C_4H^-$, $C_6H^-$ &  $C_2H$, $C_4H$, $C_6H$ \\
 & $C_8H^-$, $C_{10}H^-$, $C_{12}H^-$ &  $C_8H$, $C_{10}H$, $C_{12}H$ \\
\hline
\end{tabular}
\label{table:species}
\end{table}

\subsection{Electron transport}
 
At the gas pressure of 75 mTorr the electron mean free path is about 1 cm. This value is comparable with the width of the electrode sheath. In this case the electron energy distribution function is not determined by the local electric field. Therefore for accurate calculation of the EEDF we solve the kinetic equation for electron motion. The electron distribution function $f_e(t,x,\vec v)$ is calculated from the Boltzmann equation
\begin{equation}  \label{equation:kine}
\frac {\partial f_e}{\partial t}+ \vec v_e\frac {\partial
f_e}{\partial x}
-\frac {e\vec E}{m_e}\frac {\partial f_e}{\partial \vec v_e}=
J_e \; ,
\end{equation}
where $\vec v_e$, $m_e$ are the velocity and mass of electrons,
$E$ is the electrical field,
$J_e$ is the collisional integral  for electrons,
which includes elastic and inelastic electron scattering with argon atoms, acetylene and
other hydrocarbons molecules.

From the EEDF we can calculate the electron density by integration
\begin{equation}  \label{equation:edens}
n_e(x,t) = \int f_e d^3 v \; .
\end{equation}

To solve the Boltzmann equation we use a one-dimensional in space and three-dimensional in velocity space (1D3V) Particle in cell Monte Carlo collision (PIC-MCC) method, described in \cite{10}.

\subsection{Electron collisions}

In an argon discharge plasma containing hydrocarbons the electron-neutral collisions are the main source of reactive radicals and ions. Since the plasmochemical reactions lead to the formation of many types of species, there are also a lot of electron-neutral collision types. Nevertheless most of species have a small concentration in the background gas and we neglect some processes involving them. In fact, for modeling a mixture (5.8\% of acetylene in the inlet) only electron collisions with Ar and C$_2$H$_2$ are important for the formation of the EEDF, but we also take into account some collisions with molecular hydrogen and with heavy hydrocarbon neutrals since these processes affect the chemical balance.

We also neglect electron-neutral scattering with energy threshold much larger than the argon ionization energy since the argon ionization is the most probable inelastic process due to the large cross section and abundance of argon in the mixture. For example we neglect dissociative ionization of C$_2$H$_2$.
Table \ref{table:el_proc} shows the inelastic electron collisions taken into account in our model.

\begin{table}[ht]
\caption{the different inelastic electron-neutral collisions included in our simulation}
\centering
\begin{tabular}{l p{5.5cm} p{1.9cm} l}
\hline \hline
& Reaction & Type & Ref. \\ [0.5ex]
\hline
1 & $Ar + e \xrightarrow{}  Ar^+ + 2 e$ & Ionization & \cite{11}\\
2 & $C_{2n}H_2 + e \xrightarrow{}  C_{2n}H_2^+ + 2 e, \; n=1..4$ & Ionization & \cite{12}, \cite{21}\\
3 & $H_2 + e \xrightarrow{}  H_2^+ + 2 e$ & Ionization & \cite{13}\\
\\
4 & $C_{2n}H_2 + e \xrightarrow{}  C_{2n}H + H + e, \; n=1..4$ & Dissociation & \cite{12}\\
5 & $H_2 + e \xrightarrow{}  H + H + e$ & Dissociation & \cite{14}\\
\\
6 & $C_2H_2 + e \xrightarrow{} C_2H^- + H$ & Dissociative attachment & \cite{15}\\
\\
7 & $C_{2n}H_m^+ + e \xrightarrow{} C_{2n_1}H_{m_1} + C_{2n_2}H_{m_2}, \newline n=1..6, \; n_1+n_2=n, \; m_1+m_2=m$ & Dissociative recombination & \cite{16}\\
\\
8 & $C_2H_2^{(v=0)} + e \xrightarrow{} C_2H_2^{(v=2,3,5)} + e$ & Vibrational & \cite{17}\\
9 & $H_2^{(v=0)} + e \xrightarrow{}  H_2^{(v=1,2,3)} + e$ & excitation & \cite{18}\\
\\
10 & $Ar + e \xrightarrow{} Ar^* +e$ & Electron & \cite{19}\\
11 & $C_2H_2 + e \xrightarrow{} C_2H_2^* + e$ & level excitation & \cite{20}\\

\hline
\end{tabular}
\label{table:el_proc}
\end{table}

The cross sections for ionization of heavy hydrocarbons taken from \cite{21}, are calculated using the Binary-Encounter-Bethe model, since there is no experimental data concerning their cross sections in the literature. Dissociation processes do not affect the charge distribution directly, they only affect the chemical balance between hydrocarbons which participate in the ion production. In our simulation the dissociation cross sections of heavy hydrocarbons were taken to be the same as for acetylene due to the lack of experimental data.

We are interested in studying the electronegativity of the mixture since a large portion of Ar could make this mixture fully electropositive in contrast to the pure acetylene case.
Note that the attachment process is very important as it is the only source of negative ions in the system. We also took into account dissociative recombination of hydrocarbon cations with electrons.

The electron elastic collisions with argon atoms \cite{19}, acetylene and heavy hydrocarbons molecules \cite{20} were also considered. The hydrocarbon cross sections were taken to be equal to the one for acetylene). 
We do not take into account electron-electron collisions because the plasma density is smaller than $10^{10}$ cm$^{-3}$. The secondary electron emission from the electrode due to the ion bombardment is not included. Indeed, preliminary calculations of the discharge parameters in the Ar/C$_2$H$_2$ mixture showed that the secondary electrons do not make an important contribution in the plasma balance for a secondary electron emission coefficient $\gamma = 0.1$.

\subsection{Ion transport}
For the ions the mean free path is about 0.06 cm that is much less than the characteristic length of the electric field variation. Therefore for this gas pressure the assumption about the local dependence of the ion energy distribution function on the electrical field is quite appropriate. In this case we can use the fluid approach to calculate the ion transport. %This approach is much faster than one used for electrons, moreover kinetic approach is not an option since we have lots of ion types.
We consider many different types of negative and positive ions and the use of the fluid approach helps to accelerate the simulations and allows us to calculate the chemical processes.
Nevertheless below we will apply the kinetic approach to validate our hybrid model, (section \ref{section:model_validation}).
 
The ion transport equations 
for positive ions are
\begin{equation} \label{equation:fluid1}
\frac {\partial n_i}{\partial t} + \frac{\partial n_i v_i}{\partial x} = \nu_{ion} n_{n} n_e - \beta n_i n_e \; ,
\end{equation}
\begin{equation} \label{equation:fluid2} 
\frac {\partial n_i v_i}{\partial t} + v_i \frac{\partial n_i v_i}{\partial x} - \frac{e E}{m_i}n_i = n_i v_i \nu_{mom,i} \; ,
\end{equation}
and for negative ions
\begin{equation}  \label{equation:fluid3}
\frac {\partial n_i}{\partial t} + \frac{\partial n_i v_i}{\partial x} = \nu_{att} n_{n} n_e \; ,
\end{equation}
\begin{equation}  \label{equation:fluid4} 
\frac {\partial n_i v_i}{\partial t} + v_i \frac{\partial n_i v_i}{\partial x} + \frac{e E}{m_i}n_i = n_i v_i \nu_{mom,i} \; ,
\end{equation}
where $n_i$, $v_i$, $m_i$ are the ion concentration, velocity and mass, $n_n$ is the concentration of neutrals corresponding to the i-th ion type, $n_e$ is the electron density,
$E$ is the electrical field, $\beta$ is the dissociative recombination coefficient, $\nu_{ion}$, $\nu_{att}$ and $\nu_{mom}$ are the ionization, attachment and effective momentum transfer frequencies, respectively. In Eqs. (\ref{equation:fluid1}), (\ref{equation:fluid3}) we omitted the chemical balance terms which will be discussed in section \ref{section:chemistry}.

We do not consider radiative recombination because: a) dissociative recombination is much faster for molecular ions, and b) in the case of atomic argon ions the characteristic time of radiative recombination is much larger than for charge exchange processes.
The electrodes are considered fully absorptive for ions. 

In Eqs. (\ref{equation:fluid2}), (\ref{equation:fluid4}) we should define the effective momentum transfer frequency for each ion.
The following model was used for the description of the ion-neutral collisions.
We take into account the elastic collisions and resonance charge exchange for Ar$^+$ ions since argon is the most abundant background gas. For the other types of ions we consider only ion-neutral elastic scattering.

The mean ion energy distribution for Ar$^+$ calculated from the kinetic model (section \ref{section:model_validation}) demonstrates a significant variation so we should take into account the energy dependence of the collision cross section. When the ion energy is small, the ions interact with neutrals due to neutral polarization. The cross section of this process decreases inversely proportional to the mean relative velocity between a neutral and an ion
\begin{equation}  \label{equation:sigma1} 
\sigma_{tr} = 2\pi\sqrt{\alpha \epsilon^2 / \varepsilon^{'}} \; .
\end{equation}
Here $\alpha$ is the polarization of a neutral and $\varepsilon^{'}$ is the kinetic energy of relative motion.

At high ion energies the ion and a neutral interact like hard spheres. The cross section of this process is
\begin{equation}  \label{equation:sigma2} 
\sigma_{tr} = \pi\left(\frac{\sigma_{i}+\sigma_{k}}{2}\right)^{2} \; ,
\end{equation}
where $\sigma_{i}$ and $\sigma_{k}$ are the collision diameters of an ion and a neutral, respectively.

The resulting elastic collision cross section is the sum of these two cross sections.
To define effective cross sections for all ions in the mixture we need polarizabilities of most abundant molecules and also collision diameters for these molecules and for all ions. The required parameters for these gases are given in Table \ref{table:polar}. Effective diameters for ions were estimated based on the diameters for appropriate neutrals.

\begin{table}[ht]
\caption{The Lennard-Jones parameters ($\sigma, \epsilon$) and polarizations $\alpha$ used in our simulation}
\centering
\begin{tabular}{c c c c}
\hline \hline
Molecule & $\sigma (A)$ & $\epsilon / k_B (K)$ & $\alpha / a_0^3$  \\ [0.5ex]
\hline
$Ar$ & 3.54 & 93.3 & 11.1\\
$C_2H_2$ & 4.033 & 231.8 & 23.55 \\
$H_2$ & 2.827 & 59.7 & 5.52 \\
$C_4H_2$ & 4.5 & 320 &  \\
$C_6H_2$ & 5.05 & 397 &  \\
$C_8H_2$ & 5.9 & 470 &  \\
\hline
\end{tabular}
\label{table:polar}
\end{table}

\begin{figure}[h]
\includegraphics[width=1.0\linewidth]{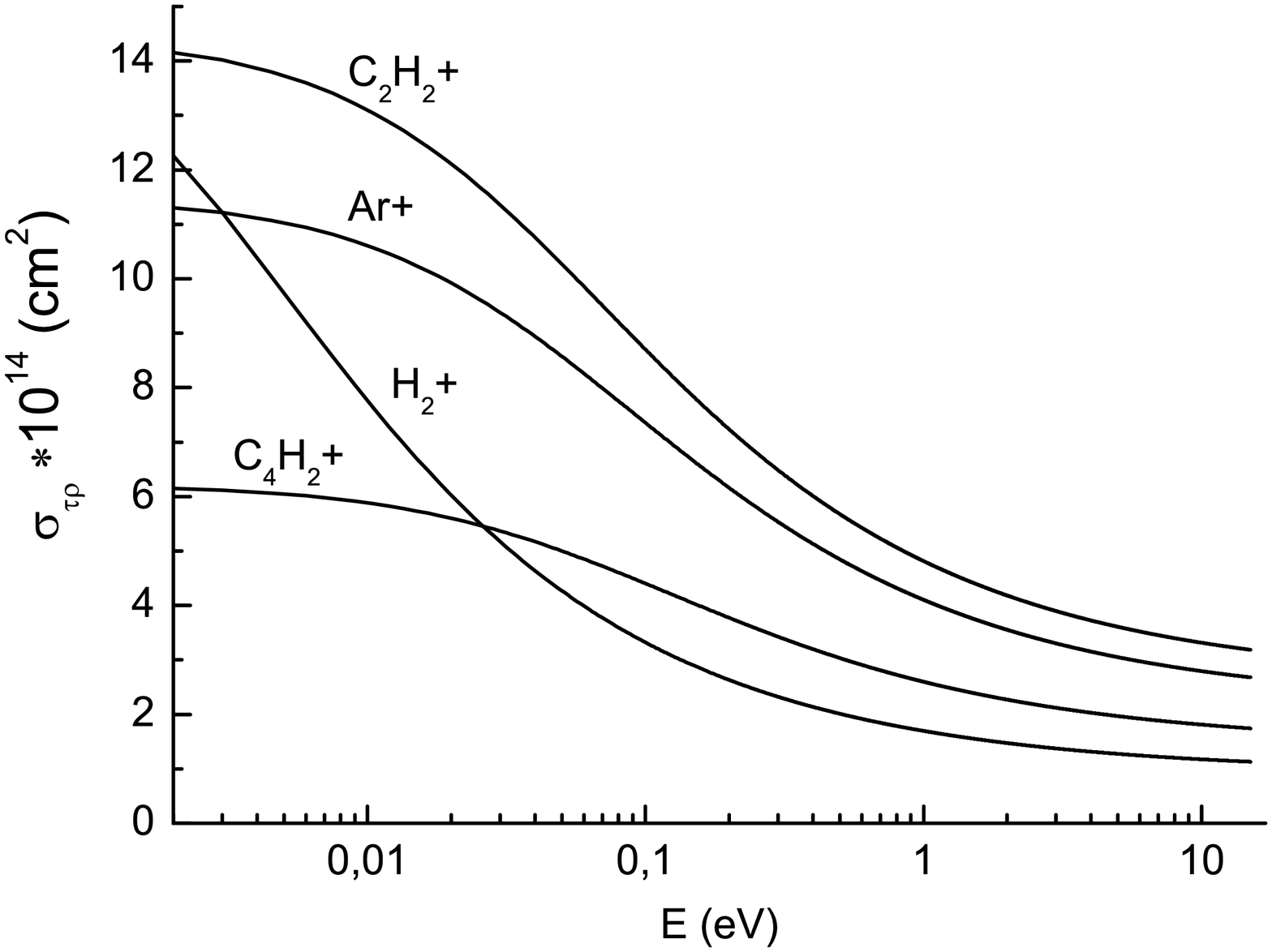}
\caption{Ion-neutral collision cross sections for different ions in a mixture of 5.8\% acetylene and argon.}
\label{figure:ar+cross}
\end{figure}

The collision cross sections for several ions calculated using this model are shown in Fig \ref{figure:ar+cross}. With these cross sections we calculate the momentum transfer frequencies
\begin{equation}  \label{equation:kin_coeff}
\nu_{mom,i}=\sum n_k v\acute{}\left( \varepsilon_i \right) \sigma_{tr,k} \left( \varepsilon_i \right) \; ,
\end{equation}
where $v\acute{}$ is the mean relative ion-neutral velocity, $n_k$ is the concentration of the k-th component of the background gas and $\sigma_{tr, k}$ is the transport cross section of the ion-neutral collision for the k-th component of the background gas.

\subsection{Poisson equation}

The Poisson equation for the electrical potential $\phi$ distribution
\begin{equation}  \label{equation:Poisson}
\bigtriangleup \phi =4\pi e \left(n_e-\sum_{i=1}^N n_i \right),
\quad
E=-\frac{\partial \phi}{\partial x} \; ,
\end{equation}
 is solved self-consistently with Eqs. (\ref{equation:kine})-(\ref{equation:fluid4}) by the iteration method.
The boundary conditions for the Poisson equation
are $U(d,t)= U_0 cos(\omega t)$ and $U(0,t)=0$, where $U_0$
 is the applied voltage amplitude and x=0, x=d
are the coordinates of the electrodes.

\subsection{Validation of the ion transport model}
\label{section:model_validation}
To check the accuracy of our hybrid approach for low gas pressure we compared the results obtained with two different models. The first one is a fully kinetic model which includes the Boltzmann equations for electrons and for one effective ion type. The second one is our hybrid model with the same type of ions. We consider only the ion-neutral elastic scattering.

The parameters of the discharge were taken the same as in the experiment: the interelectrode distance is 7 cm, the gas pressure is 75 mTorr, the frequency and amplitude of the applied voltage are 13.56 MHz and 92 V, respectively.

\begin{figure}[h]
\includegraphics[width=1.0\linewidth]{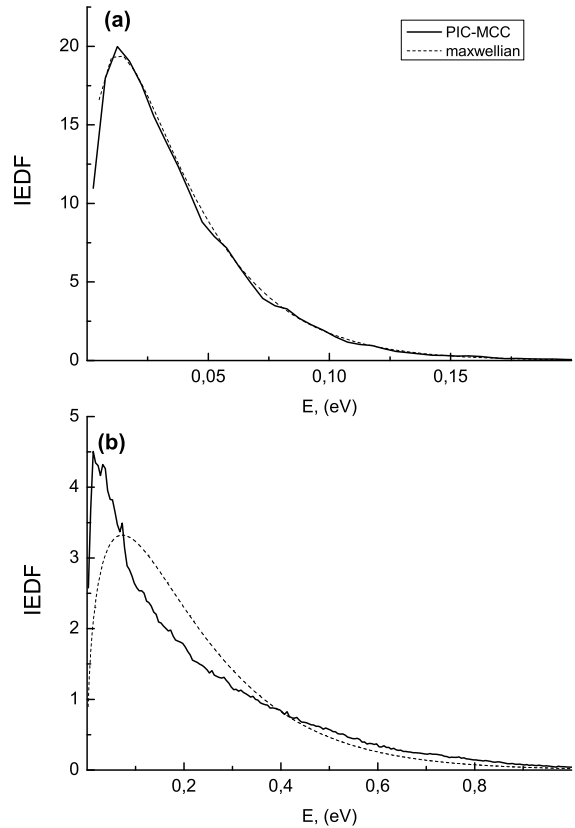}
\caption{Ion energy distribution function in the middle of discharge gap (x=3.5 cm) (a) and in the sheath region (x=0.75 cm) (b).}
\label{figure:iedf11}
\end{figure}

Since the elastic scattering cross section is energy dependent, the  momentum transfer frequency is dependent on the local ion energy distribution function (IEDF). In Fig. \ref{figure:iedf11} the IEDFs calculated with the full kinetic model are shown at different coordinates. For our conditions the IEDF appeared to have an almost Maxwellian distribution in the bulk plasma and is very different in the sheath region. Nevertheless we can consider the Maxwellian distribution as a good approximation to determine the ion kinetic coefficients.

\begin{figure}[h]
\includegraphics[width=1.0\linewidth]{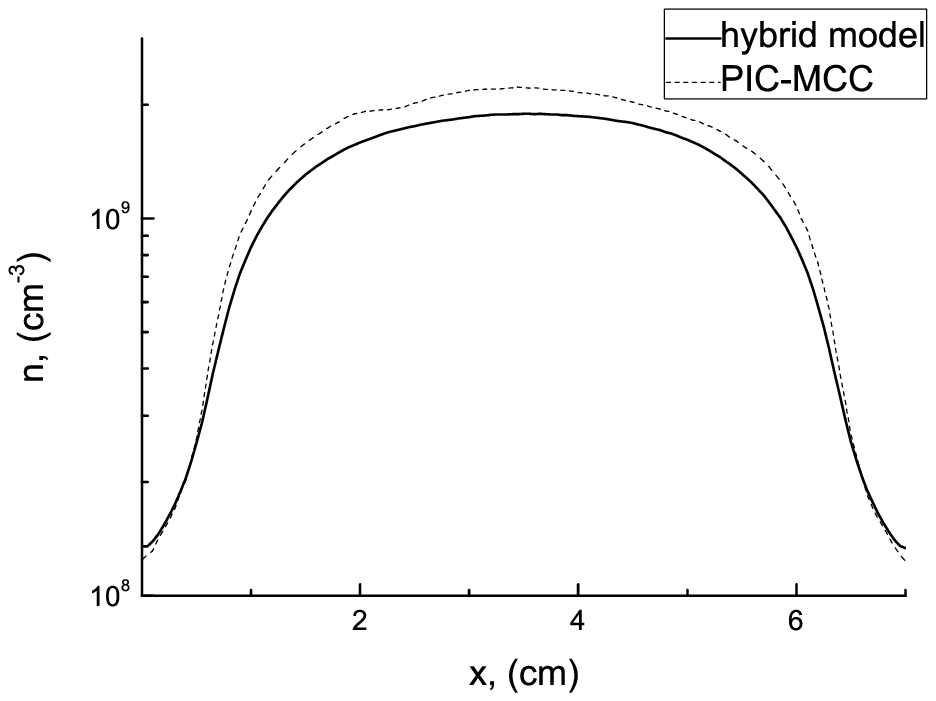}
\caption{Ion density distribution obtained with the hybrid model (solid line) and the PIC-MCC simulations (dashed line).}
\label{figure:comp1}
\includegraphics[width=1.0\linewidth]{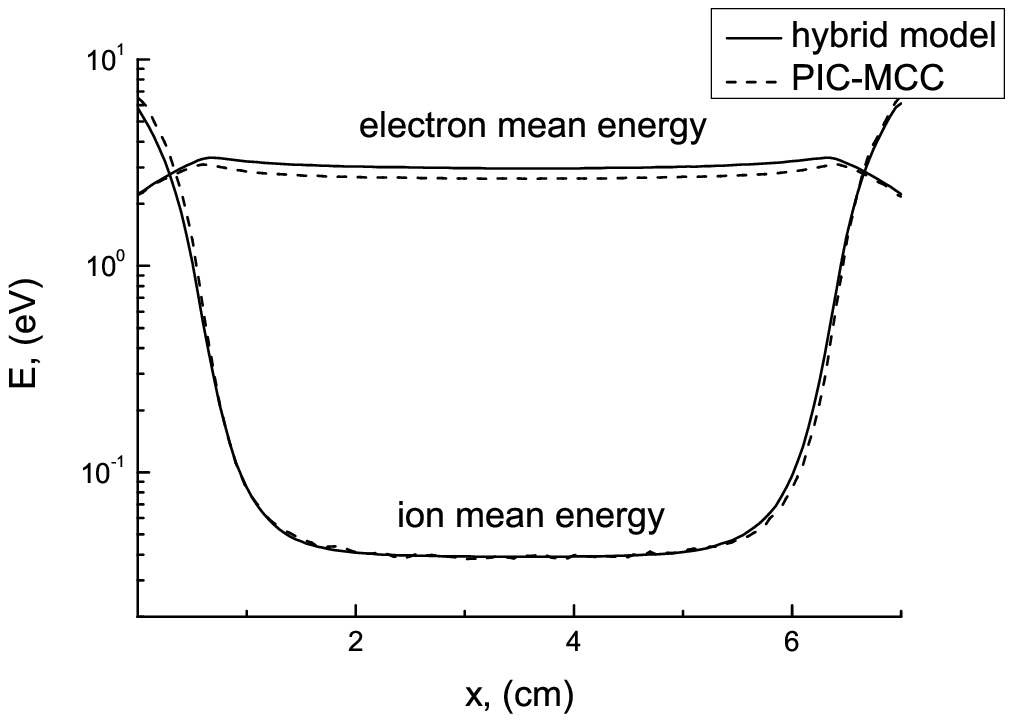}
\caption{Electron and ion mean energy distribution obtained with the hybrid model (solid line) and the PIC-MCC simulations (dashed line).}
\label{figure:comp2}
\end{figure}

Fig. \ref {figure:comp1} and \ref {figure:comp2} show the ion density, ion and electron energies obtained using the hybrid model and in the PIC-MCC simulations. They demonstrate a good agreement between the two models for the ion parameters distribution. The electron EDF is a sensitive parameter and it was observed that the mean electron energy decreases in the case of a fully kinetic simulation. But anyway this deviation is about 15\% which is close to the accuracy of a kinetic model.

\subsection{Neutral transport}
We consider the following balance equations for the neutral density distribution
\begin{equation}  \label{equation:fluidn}
\frac {\partial n_n}{\partial t} - \nabla(D_n \nabla( n_n )) = \beta n_i n_e - \nu_{ion} n_{n} n_e + S_{react,j} \; ,
\end{equation}
where $D_n$ is the diffusion coefficient, $S_{react,j}$ is a source term which represents flow, pump and mixing processes. Diffusion coefficients for different species in the mixture are obtained from pair diffusion coefficients using Blanc formula
\begin{equation}  \label{equation:diffn}
\frac {P_{tot}}{D_n} =  \sum \left(\frac {P_j}{D_{n,j}}\right) \; ,
\end{equation}
where $P_{tot}$ is the gas pressure, $P_j$ is the pressure of the j-th component of the background gas and $D_{n,j}$ is the pair diffusion coefficient.
These pair diffusion coefficients can be obtained from Lennard-Jones parameters of neutral molecules using Chapman-Enskog theory \cite{24,25}.
For two species with masses $m_i$ and $m_j$, and Lennard-Jones parameters ($\sigma_i$, $\varepsilon_i$) and
($\sigma_j$, $\varepsilon_j$), the binary diffusion coefficient at the  pressure $P$ and temperature $T$ is given by
\begin{equation}  \label{equation:diffnpair}
D_{i,j} =  \frac{3}{16} \frac{k_B T}{P} \frac{2 \pi k_B T/m_{i,j}}{\pi \sigma^2_{i,j} \Omega_D(\Psi)} \; ,
\end{equation}
where $m_{i,j} = m_i m_j/(m_i + m_j)$ is the reduced mass,
\begin{equation}  \label{equation:diffnsigma}
\sigma_{i,j} = \frac{\sigma_{i} + \sigma_{j}}{2} \; ,
\end{equation}
is the binary collision diameter and $\Psi = T/\sqrt{\epsilon_i \epsilon_j}$.
$\Omega_D(\Psi)$ is calculated from the expression \cite{25}
\begin{equation}  \label{equation:diffnomega}
\Omega_D(\Psi) = \frac{A}{\Psi^B} + \frac{C}{e^{D \Psi}} + \frac{E}{e^{F \Psi}} + \frac{G}{e^{H \Psi}} \; ,
\end{equation}
with $A = 1.06036, B = 0.15610, C = 0.19300, D = 0.47635, E = 1.03587, F = 1.52996,
G = 1.76474, H = 3.89411$.
The required parameters for different neutrals were estimated based on the values given in Table \ref{table:polar}.

$S_{react,j}$ can be presented as a sum
\begin{equation}  \label{equation:pump}
S_{react,j} = S_{flow,j} + S_{pump,j} + S_{mix,j} \; ,
\end{equation}
where $S_{flow,j}$ represents the gas inlet, $S_{pump,j}$ is the pumping source term and $S_{mix,j}$ is the mixing source term.
We use the same model as in \cite{8} and the assumption of a perfectly stirred reactor. It gives the following expression
\begin{equation}  \label{equation:pump2}
S_{pump,j} = - \frac{n_n}{\tau} \; ,
\end{equation}
where $\tau$ is the average residence time of all neutrals. The $\tau$ is adjusted in such a way that the pressure in the discharge equals the desired pressure.

\begin{equation}  \label{equation:pump3}
S_{mix,j} = \left(S_{flow,j} + S_{pump,j} \right) \left( \frac{V_{react} - V_{disch}}{V_{disch}}\right) \; ,
\end{equation}
where $V_{react}$ is the total volume of the plasma reactor and
$V_{disch}$ is the discharge volume.

The electrodes are considered as fully reflective for molecules. For radicals we adopted a sticking model described in \cite{8}, with sticking coefficients taken from \cite{9}. We do not take into account wall reactions.

\subsection{Chemical processes}
\label{section:chemistry}
Besides the few ions and radicals which are generated by electron-molecule collisions all other species appear as a result of chemical processes.
As a base for modeling chemical processes we used the set of reactions proposed in \cite{9} where a detailed description of the model and appropriate references can be found. The further development of the reaction set is presented in \cite{Ming}.
The considered reactions are shown in Table \ref{table:reactions}. Note that, the chemical balance terms are included in the transport equations (\ref{equation:fluid1}), (\ref{equation:fluid3}), (\ref{equation:fluidn}).

\begin{table}[h]
\caption{The chemical reactions used in our simulation}
\centering
\begin{tabular}{l}
\hline \hline
Reaction \\ [0.5ex]
\hline
 Cluster growth through hydrocarbon cations\\
 $C_2H^+ + H_2 \rightarrow C_2H_2^+ + H$\\
 $C_2H^+ + C_2H_2 \rightarrow C_4H_2^+ + H$\\
 $C_2H_2^+ + C_2H_2 \rightarrow C_4H_2^+ + H_2$\\
 $C_2H_2^+ + C_2H_2 \rightarrow C_4H_3^+ + H$\\
 $C_4H_2^+ + C_2H_2 \rightarrow C_6H_4^+$\\
 $C_4H_3^+ + C_2H_2 \rightarrow C_6H_5^+$\\
 $C_6H_2^+ + C_2H_2 \rightarrow C_8H_4^+$\\
 $C_6H_4^+ + C_2H_2 \rightarrow C_8H_6^+$\\
 $C_8H_4^+ + C_2H_2 \rightarrow C_{10}H_6^+$\\
 $C_8H_6^+ + C_2H_2 \rightarrow C_{10}H_6^+ + H_2$ \\
 $C_{10}H_6^+ + C_2H_2 \rightarrow C_{12}H_6^+ + H_2$ \\
 \\
 Cluster growth through hydrocarbon anions \\
 $C_{2n}H^- + C_2H_2 \rightarrow C_{2n+2}H^- + H_2, \; n=1..5$\\
 \\
 Cluster growth through $C_2H$ insertion\\
 $C_2H + H_2 \rightarrow C_2H_2 + H$\\
 $C_2H + H \rightarrow C_2H_2$\\
 $C_2H + C_{2n}H_2 \rightarrow C_{2n + 2}H_2 + H, \; n=1..5$\\
 \\
 Cluster growth through acetylene insertion\\
 $C_{2n}H + C_2H_2 \rightarrow C_{2n +2}H_2 + H, \; n=2..5$\\
 \\
 Neutralization reactions\\
 $C_{2n_1}H^- + C_{2n_2}H_m^+ \rightarrow C_{2n_1}H + C_{2n_2}H_m, \; n_1,n_2=1..6$\\
 $C_{2n_1}H^- + Ar^+ \rightarrow C_{2n}H + Ar$\\
 $C_{2n_1}H^- + ArH^+ \rightarrow C_{2n}H + Ar + H$\\
 $C_{2n_1}H^- + H_2^+ \rightarrow C_{2n}H + H + H$\\
 \\
 Charge exchange reactions\\
 $H_2^+ + C_2H_2 \rightarrow H_2 + C_2H_2^+$\\
 $Ar^+ + C_2H_2 \rightarrow Ar + C_2H_2^+ \;$ \cite{27}\\
 $Ar^+ + H_2 \rightarrow Ar + H_2^+ \;$\cite{27}\\
 \\
 Hydrogen insertion and hydrogen abstraction\\
 $H + C_{2n}H_2 \rightarrow C_{2n}H_3, \; n=1..3$\\
 $H + C_{2n}H_3 \rightarrow C_{2n}H_2  + H_2, \; n=1..3$\\
 $H + C_{2n}H \rightarrow C_{2n}H_2, \; n=2..6$\\
 $H_2 + C_{2n}H \rightarrow C_{2n}H_2  + H, \; n=2..6$\\
 \\
 Other reactions\\
 $Ar^+ + H_2 \rightarrow ArH^+ + H \;$ \cite{27}\\
 $C_2H + C_2H_3 \rightarrow C_2H_2 + C_2H_2$\\
 $C_2H_2 + C_2H \rightarrow C_4H_3$\\
 $C_4H_2 + C_2H \rightarrow C_6H_3$\\
 $C_4H_3 + H \rightarrow C_2H_2 + C_2H_2$\\
 $C_6H_3 + H \rightarrow C_4H_2 + C_2H_2$\\
% \hline{Other}
 
\hline
\label{table:reactions}
\end{tabular}
\end{table}

The main difference with respect to the reaction set considered for pure acetylene in \cite{9} is the addition of argon involving charge exchange reactions \cite{27}.
Due to these reactions a large part of argon ions is converted to hydrocarbon and hydrogen ions. The argon presence can be considered as an additional source of positive ions. Also due to recombination reactions it is a sink for negative ions.
Thus even if argon ions are not the most abundant, the presence of argon as a background gas makes the mixture more electropositive.

\section{Results and analysis}
\label{section:res_desc}
First we analyze how the presence of a small fraction of acetylene and the related chemical processes affect discharge properties.

\subsection{Acetylene influence}
\label{section:acet_influence}
Let us evaluate the influence of 5.8\% acetylene in argon on the discharge properties for the case when acetylene is just injected in the discharge and heavy hydrocarbons are not formed yet. So we excluded all chemical reactions except neutralization from the chemical balance.

\begin{figure}[h]
\includegraphics[width=1.0\linewidth]{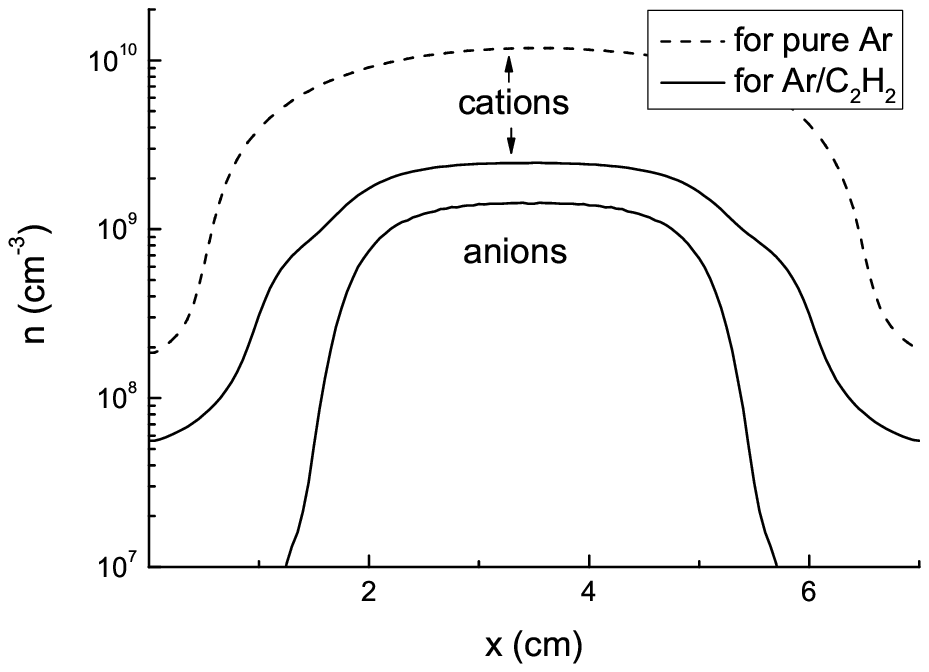}
\caption{Ion density distributions for pure argon and for argon with 5.8\% of acetylene.}
\label{figure:acet_compare}
\includegraphics[width=1.0\linewidth]{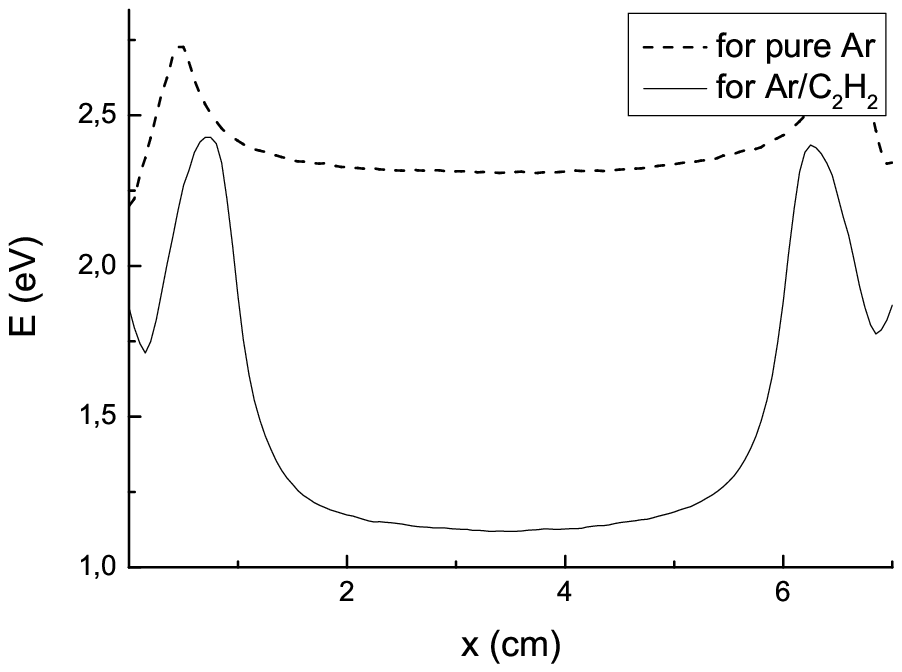}
\caption{Mean electron energy distributions for pure argon and for argon with 5.8\% of acetylene.}
\label{figure:acet_compare_ene}
\end{figure}

\begin{figure}[h]
\includegraphics[width=1.0\linewidth]{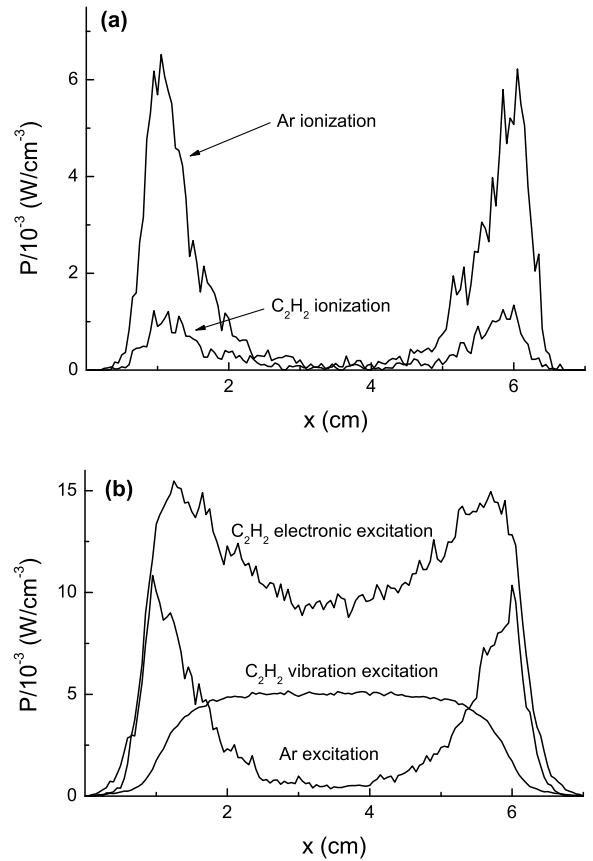}
\caption{Power consumed by ionization processes (a) and different excitation processes (b) for argon with 5.8\% of acetylene.}
\label{figure:ionization_consum}
\end{figure}

%It was observed that
Acetylene addition causes a decrease of the positive ion density. In Fig. \ref{figure:acet_compare} the ion density distributions are shown for pure argon and with 5.8\% addition of acetylene. In the second case the density of positive ions is almost 5 times lower. It can be explained by two reasons. The most important one is the change of the EEDF and the other one is the ion-ion recombination. Indeed, in the case of acetylene addition negative ions are formed in the plasma, which can recombine with the positive ions. In Fig. \ref{figure:acet_compare_ene} the mean electron energy distribution is shown for pure argon and for the mixture. In the case when acetylene is present in the  discharge the mean electron energy is lower (almost twice in the bulk plasma). This is explained by the redistribution of the discharge power between the collision processes. In Fig. \ref{figure:ionization_consum} the power consumed by ionization and excitation processes are shown. It is seen than even a small concentration of acetylene consumes a large portion of the discharge power for excitation especially in the bulk region. The reason is that the vibrational excitation threshold of C$_2$H$_2$ is small and these inelastic collisions efficiently cool the electrons in the midplane,  decreasing in this way the mean electron energy and suppressing ionization.

The presence of acetylene also changes the electronegativity of the mixture. We found that the density of negative ions reaches 50\% with respect to the positive ion density even for such small content of acetylene in the background gas (Fig. \ref{figure:acet_compare}). The density of negative ions is considerable, although the electron impact ionization and charge exchange reactions lead to the creation of much larger quantity of positive ions. However, the negative ions are trapped in the bulk plasma due to the electrical potential distribution and can rarely reach the electrode. In such conditions the main sink for anions is recombination which is rather slow. As a result the anions can reach sufficient concentrations in the middle of the discharge.

\subsection{Cluster growth influence}
\begin{figure}[h]
\includegraphics[width=1.0\linewidth]{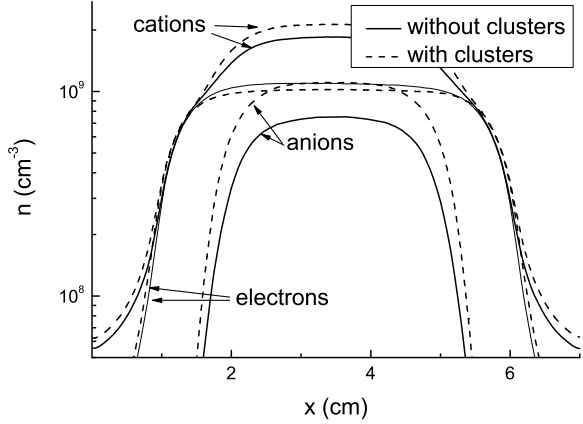}
\caption{Total (positive and negative) ion and electron density distributions without hydrocarbon clusters(initial stage) and with them.}
\label{figure:cluster_compare}
\end{figure}

During cluster growth the light cations and anions convert into heavy ones. The concentration of heavy cations increases due to their lower mobility and small recombination rate with anions.
Due to the latter reason the concentration of heavy anions also increases. In Fig. \ref{figure:cluster_compare} the electron and ion density distributions are shown for the case with hydrocarbon clusters with maximum 12 carbon atoms and for the case in the absence of clusters (i.e., the initial stage, before cluster growth starts taking place). It is seen that the electron density distribution over the discharge gap changes just a little. Also the electron mean energy distribution practically does not change (less than 5 $\%$) with cluster growth. It is related to the fact that the positive and negative ions densities increase simultaneously and, therefore, the spatial charge distribution changes insignificantly.

\begin{figure}[h]
\includegraphics[width=1.0\linewidth]{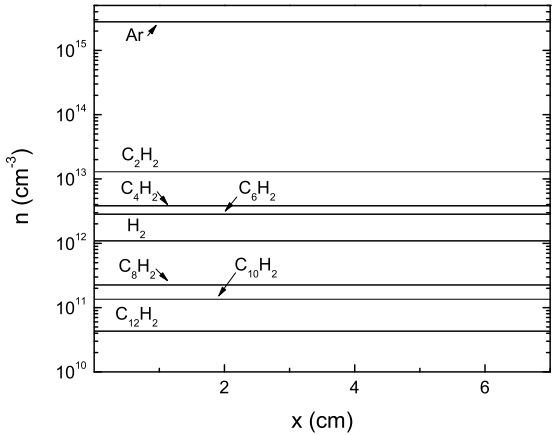}
\caption{Densities of the various neutral molecules.}
\label{figure:neutrals}
\end{figure}

Another important sequence of cluster growth and low gas pumping speed is the conversion of acetylene to heavy neutrals $C_{2n}H_2$ at much larger time scale. It finally leads to a significant neutrals concentration variation i.e., a decreasing concentration of acetylene and an increase of heavy hydrocarbons and argon concentration. This observation is in good agreement with experiments \cite{6,7} where the authors observed significant acetylene monomer dilution. The resulting neutrals concentrations are shown in Fig. \ref{figure:neutrals}. The calculated acetylene concentration is almost ten times less than the initial 5.8\% of the total concentration, as a result of cluster growth and deposition to the walls.
This conversion makes electron collisions with such hydrocarbons like C$_4$H$_2$, C$_6$H$_2$ more important. Since we use approximate cross sections for the ionization of C$_4$H$_2$, C$_6$H$_2$ molecules, it makes our results less reliable if their density increases.

\subsection{Species density distribution}
\begin{figure}[h]
\includegraphics[width=1.0\linewidth]{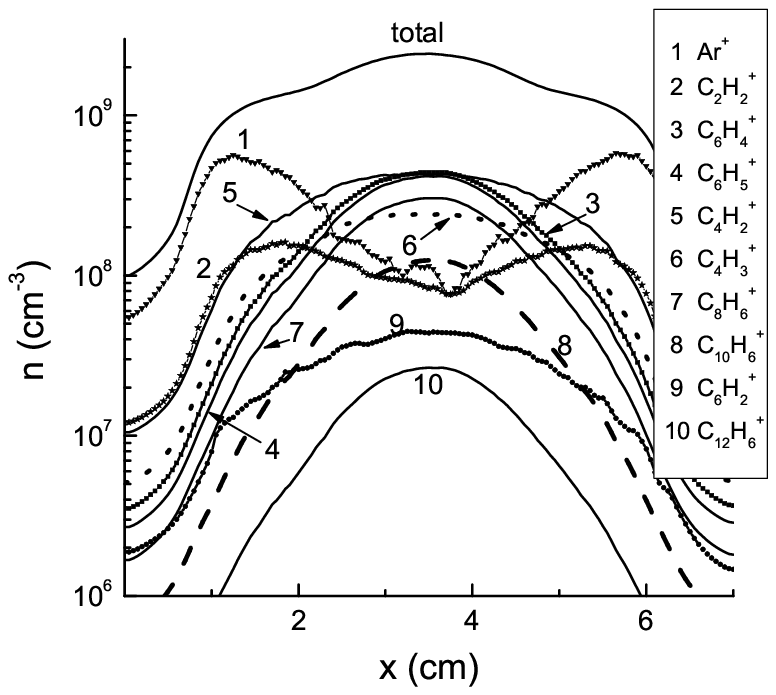}
\caption{Density distributions of the various cations.}
\label{figure:posions}
\end{figure}

\begin{figure}[h]
\includegraphics[width=1.0\linewidth]{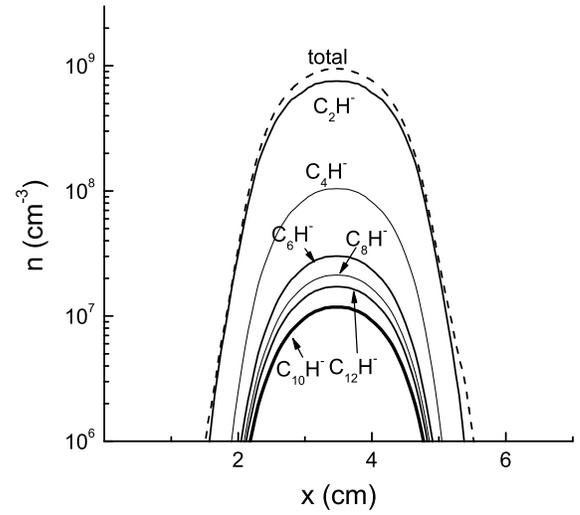}
\caption{Density distributions of the various anions.}
\label{figure:negions}
\end{figure}

\begin{figure}[h]
\includegraphics[width=1.0\linewidth]{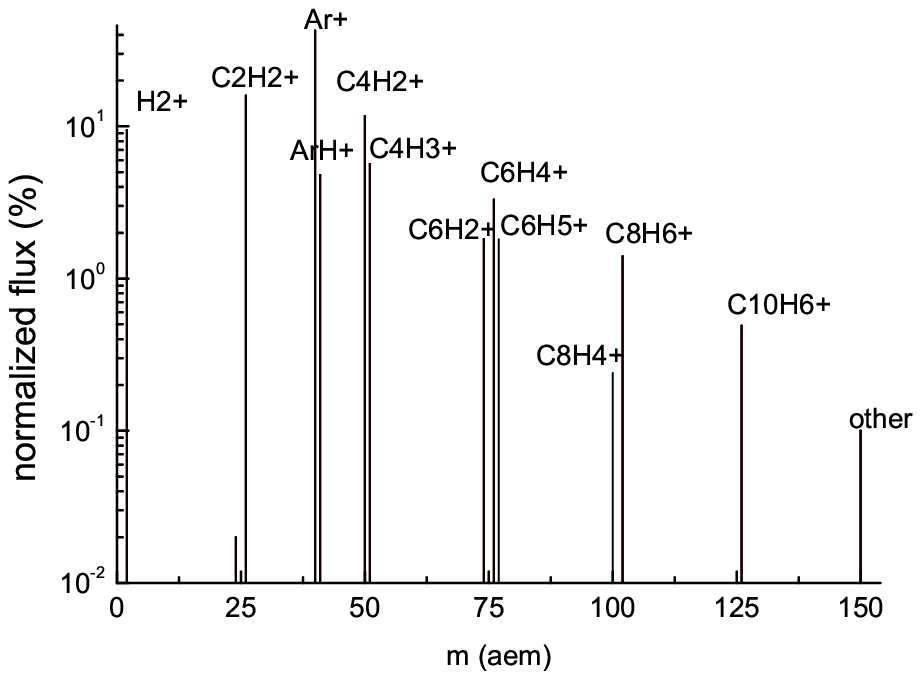}
\caption{Positive ion spectrum (normalized flux of different cations to electrode).}
\label{figure:spectra1}
\includegraphics[width=1.0\linewidth]{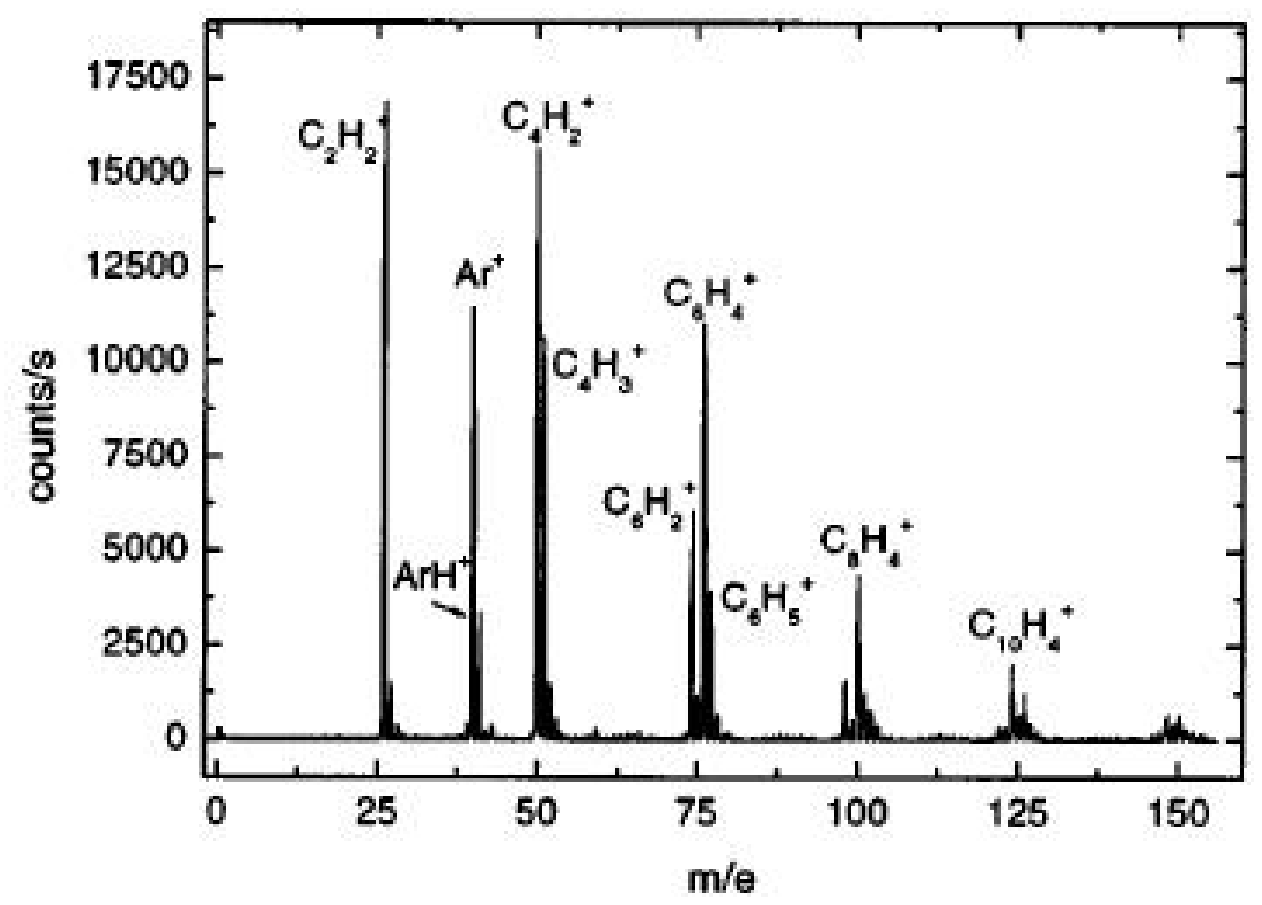}
\caption{Experimental positive ion spectrum.}
\label{figure:spectra_exp}
\includegraphics[width=1.0\linewidth]{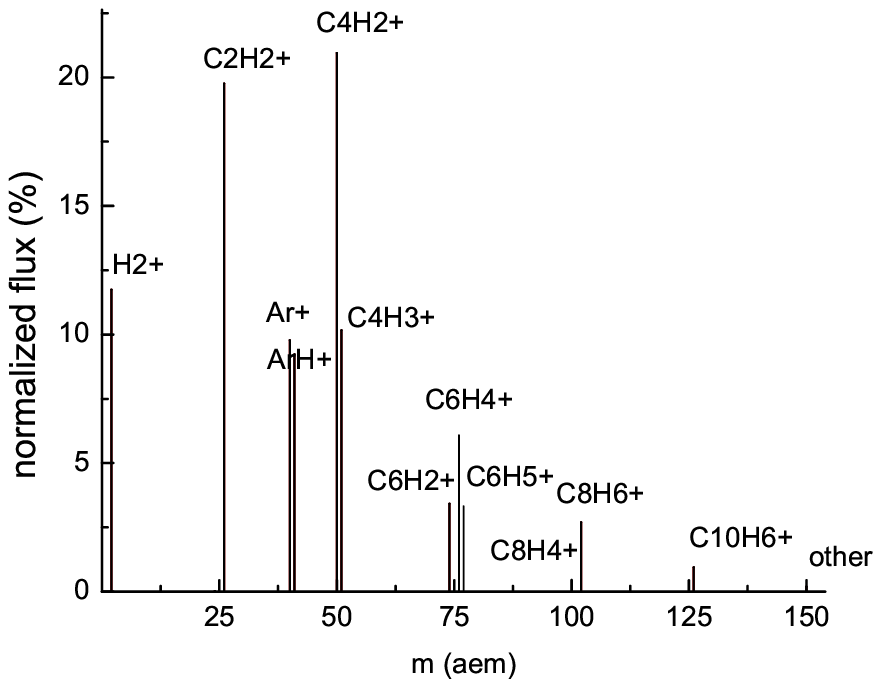}
\caption{Positive ion spectrum (only ions with energy near 10eV were taken).}
\label{figure:spectra2}
\end{figure}

As a result of the electron inelastic collisions and the plasmochemical processes we obtained the steady state ion density distribution shown in Figs. \ref{figure:posions} and \ref{figure:negions}. 
The heavy cations become abundant in the middle of the discharge due to the conversion and due to their lower mobility. Near the sheath-plasma boundary the electron impact ionization has a maximum. As a result the densities of Ar$^+$ and C$_2$H$_2^+$ ions have two peaks and the Ar$^+$ ions dominate near the sheath region. It should provide an intense argon ion flux to the electrode. We have compared the calculated ion flux to the electrode with the mass spectrum measurements from \cite{6,7}. Fig. \ref{figure:spectra1} shows the calculated normalized ion flux of different ions to the electrode presented as a function of the ion mass. In Fig. \ref{figure:spectra_exp} the measured \cite{6,7} positive ion spectrum is shown. The experimental spectrum does not demonstrate a significant argon ion flux in contrast with the calculated one. The reason is probably that in the experiments only ions are counted with energy lying in the fixed energy window near a certain energy value. Due to the varying mobility the different types of ions have considerably different mean energies near the electrode. Therefore only a part of the ions arriving at the electrode are measured. In the case of Ar$^+$ ions the mobility and consequently, the energy is rather small because in our mixture the Ar$^+$ ions participate in the charge exchange resonant collisions and in the elastic scattering. If we took the Maxwellian distribution for ions then the flux of ions with certain energy can deviate to a great extent from the total flux. Considering this we obtain the following ion fluxes to the electrode shown in Fig. \ref{figure:spectra2}. Now a much better agreement is reached with the experimental data.

The negative ions reach a sufficient concentration but in contrast with the positive ions the lightest anion C$_2$H$^-$ is the dominant one, with a density of about 80\% of all anions density. This density in the middle of the discharge is twice smaller than the cation density so we can conclude that this mixture has significant electronegativity regardless of a small concentration of acetylene and other hydrocarbons.

Heavy cations C$_{12}$H$_6^+$ and anions C$_{12}$H$^-$ are present in the mixture with large concentration (about 10$^7$ cm$^{-3}$) and can be considered as precursors for dust particle formation.

\section{Conclusion}
\label{section:con_desc}
We developed a hybrid model for simulations of the 13.56 MHz discharge in a C$_2$H$_2$/Ar mixture at a gas pressure of 75 mTorr. 
This hybrid model combines a kinetic description for electron motion and the fluid approach for 6 negative and 16 different positive ions. 146 different chemical reactions were taken into account. We consider the formation of heavy hydrocarbons up to 12 carbon atoms. Both negatively and positively charged heavy hydrocarbons  can be precursors for nanoparticles formation in the discharge volume, since their densities are sufficiently large ($\approx $10$^7$~cm$^{-3}$).
The total density of negative ions reaches about one half of the positive ion density. Thus a small fraction of acetylene (5.8$\%$) in the argon discharge makes the mixture electronegative, because the negative ions are trapped in the quasineutral plasma.
We have also found that injection of 5.8$\%$ acetylene in argon decreases the plasma density by a factor of 5. With acetylene added a large part of the discharge power is transferred into electron excitation of   C$_2$H$_2$ molecules.
The cluster growth does not affect the electron density and the mean energy to a large extent, but the densities of positive and negative ions increase since the heavy ions have smaller mobility.
For the conditions of the experiments \cite {6,7} the important sequence of cluster growth and low gas pumping is a significant decrease of acetylene and an increase of heavy hydrocarbon and argon concentrations. 

\begin{acknowledgments}
This work was supported by the Bilateral grants RFBR-Flanders (05-02-19809-MFa), RFBR-Ukraine (08-02-90446-Uka).
We also acknowledge IAP program, Center of Excellence NANO and Calcua supercomputer of the University of Antwerp for supporting this work.
\end{acknowledgments}

\end{document}